# TYPING OF DATA TRANSFER PROCESSES IN THE INFORMATION SYSTEM WITHIN THE FRAMEWORK OF THREAT MODELING

*E.S. Romanova, A.K. Novokhrestov, A.A. Konev*

**Annotation:** the work is aimed at automating the process of obtaining a list of security threats aimed at the information system in the work processes of data transfer are considered, definitions for each process are presented. The typification of processes and the formalization of the list of basic data transfer processes are considered. Based on the presented typical data transmission processes, schemes of these processes have been developed that describe transmission channels and information carriers.

In this work, the process of data transfer means a set of processes in which interaction with information takes place.

In accordance with the National Standard of the Russian Federation "Information Protection. Ensuring information security in the organization. Basic terms and definitions" information process includes the process of creating, collecting, processing, accumulating, storing, searching, distributing and using information. It is also worth noting that often processes can be considered from two sides - these are processes associated with the use of electronic computers and without it. We also present process diagrams, where:

$V1$ – paper carrier of information;
$V2$ – human;
$V3$ – digital information storage device;
$V4$ – a process.

Information transfer channels:
$e1$ – in the visual environment;
$e2$ – in an acoustic environment;
$e3$ – in the electromagnetic environment;
$e4$ – in a virtual environment.

Channels for remote information transmission:
$e3`$ – in the electromagnetic environment;
$e4`$ – in a virtual environment.

Creation of information is the process of producing new information. When information is creating, the source of information appears - directly the creator of the information, the place of storage - where the created information is stored and the environment in which the information is creating. Thus, the processes of creating and storing information are directly relating to each other, because in the absence of storage of the created information, it does not make sense, namely, the basic properties of information are violating - accessibility, usefulness, completeness, accuracy, and others, since no one except the creator has access to information. Creation involves various processes, including:

Create a document on paper by hand. Channels - visual, information carrier - person, paper (Figure 1).

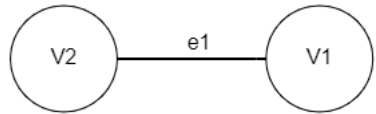

Figure 1 – Creating a paper document

Creating a document using technical means (Figure 2):
- Document on PC
- Photo / video documents using a camera, camera

Storage medium - computer hard drive, camera/camera memory card

Information transfer channels – e1 – device monitor screen, e2 – speakers, e3 – device (computer, camera, camera), e4 – computer operating system, camera or camera firmware.

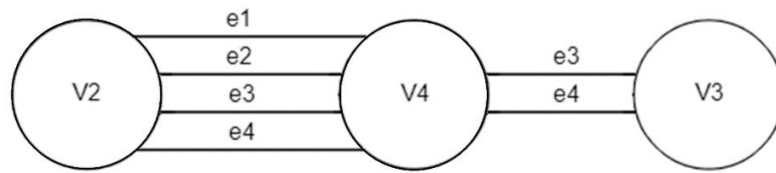

Figure 2 – Creation of an electronic document

Data collection is the process of obtaining information from the outside world and bringing it to the form standard for a given information system, the activity of the subject, during which he receives information about the object of interest to him. In addition to the already mentioned that the process can be carried out from the standpoint of the use of technical means and without their use, it is also worth noting a new type of process - "Man-man". Figures 3-6 show the main processes for collecting information.

Survey of people. Channels - acoustic (the room in which the survey takes place), the carrier of information - a person.

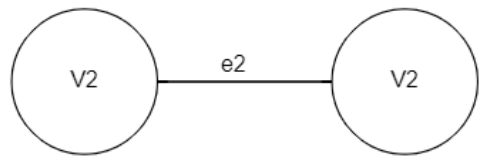

Figure 3 – Poll

Collection of information from paper sources. Channels - visual, information carrier - person, paper.

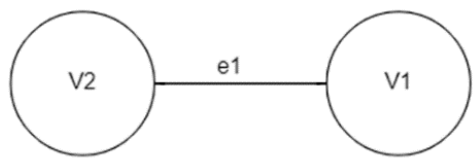

Figure 4 – Collection of information from paper sources

Local collection of information. Information transmission channels – e1 – device screen, e2 – speakers, e3 – device (computer), e4 – computer operating system. The storage medium is a computer hard disk, RAM.

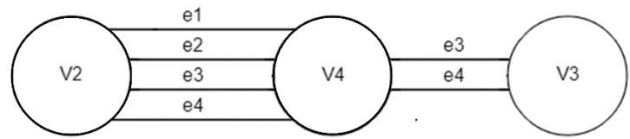

Figure 5 – Local collection of information

Remote search for information. Information transmission channels – e1 – device screen, e2 – speakers, e3 – device (computer), e4 – computer operating system, e3 – electromagnetic channels for remote data transmission, e4` – data transfer protocols. The storage medium is a computer hard disk, RAM.

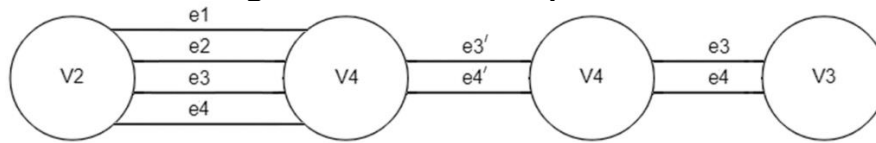

Figure 6 - Remote information search

Information search is the process of identifying in a certain set of sources all the information that belongs to a specified topic, or satisfies a predetermined search condition (request) or contains the necessary (corresponding to information needs) facts, information, data. The search is part of the collection of information, since when searching for the necessary data, we filter out what is not relating to the topic and collect the remaining information together, which, in turn, smoothly turns into the process of accumulating information.

Information processing is a purposeful process of changing the form of its presentation or content. That is, the processing is directing only to the information itself and does not contain the process of data transmission in an explicit form. Processing is directly relating to the storage of information.

Storage is the process of maintaining the original information in a form that ensures the issuance of data at the request of end users in a timely manner.

Accumulation of information - the process of forming the original, unsystematized array of information. It is worth clarifying that accumulation is the process of changing the original data, which means that it is a dynamic process, in contrast to static storage, in which information is not transforming in any way.

Information search is the process of identifying in a certain set of sources all the information that belongs to a specified topic, or satisfies a predetermined search condition (request) or contains the necessary (corresponding to information needs) facts, information, data. The search is part of the collection of information, since when searching for the necessary data, we filter out what is not relating to the topic and collect the remaining information together, which, in turn, smoothly turns into the process of accumulating information.

Distribution of information is the process of moving messages in space in the form of signals from one object to another.

The dissemination of information is very similar to the collection, but is bi-directional and in a broad sense represents the whole Human-to-Person relationship, but is dividing according to the ways people interact with each other. Figures 7-10 show typical schemes of information dissemination processes.

1. Man – process
   – Communication in instant messengers / social networks / via e-mail.

Information transmission channels – e1 – device screen, e2 – speakers, e3 – device (computer), e4 – computer operating system, e3 – electromagnetic channels for remote data transmission, e4` – data transfer protocols. The carrier of information is a person, RAM.

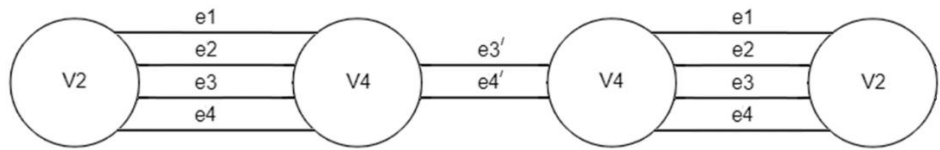

Figure 7 – Communication via the Internet

2. Man – information accumulator
   – Sending a letter (paper)

Source of information – a person, information carrier – paper, data transmission channel e1 – visual (text on paper)

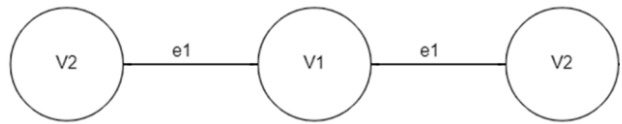

Figure 8 – Writing/reading a letter

3. Man - Man
   – Negotiations

The source of information is a person, the data transmission channel e2 is acoustic (the room in which negotiations are held)

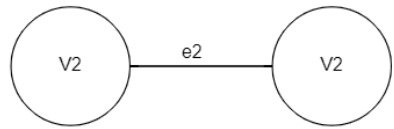

Figure 9 – Negotiations

   – Talking on the phone

Information transfer channels – e2 – phone speakers, e3 – device (phone), e4 – phone firmware, e3 – telephone communication lines, e4` – telephone protocols. The carrier of information is a person.

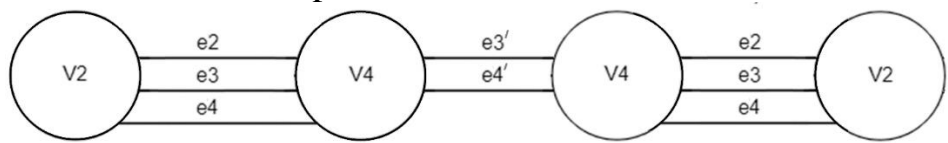

Figure 10 – Talking on the phone

This research was funded by the Ministry of Science and Higher Education of Russia, Government Order for 2020–2022, Project No. FEWM-2020-0037 (TUSUR).